\documentclass[%
 reprint,
showpacs,preprintnumbers,
 amsmath,amssymb,
 aps,
]{revtex4-1}

\usepackage{graphicx}
\usepackage{dcolumn}
\usepackage{bm}
\usepackage[colorlinks=true,linkcolor=blue,citecolor=blue,urlcolor=black]{hyperref}

\begin{document}

\title{Valley Zeeman effect and spin-valley polarized conductance in monolayer MoS$_2$ in a perpendicular magnetic field}
\date{\today}
\author{Habib Rostami}
\email{rostami@ipm.ir}
\affiliation{School of Physics, Institute for Research in Fundamental Sciences (IPM), Tehran 19395-5531, Iran}
\author{Reza Asgari}
\email{asgari@ipm.ir}
\affiliation{School of Physics, Institute for Research in Fundamental Sciences (IPM), Tehran 19395-5531, Iran}
\begin{abstract}
We study the effect of a perpendicular magnetic field on the electronic structure and charge transport of a monolayer MoS$_2$ nanoribbon at zero temperature. We particularly explore the induced valley Zeeman effect through the coupling between the magnetic field, $B$, and the orbital magnetic moment. We show that the effective two-band Hamiltonian provides a mismatch between the valley Zeeman coupling in the conduction and valence bands due to the effective mass asymmetry and it is proportional to $B^2$ similar to the diamagnetic shift of exciton binding energies. However, the dominant term which evolves with $B$ linearly, originates from the multi-orbital and multi-band structures of the system. Besides, we investigate the transport properties of the system by calculating the spin-valley resolved conductance and show that, in a low-hole doped case, the transport channels at the edges are chiral for one of the spin components. This leads to a localization of the non-chiral spin component in the presence of disorder and thus provides a spin-valley polarized transport induced by disorder.
\end{abstract}
\pacs{75.70.Ak, 72.25.-b, 73.43,-f}
\maketitle

\section{introduction}

Monolayer of the molybdenum disulfide (ML-MoS$_2$) has recently
attracted great interest because of its potential applications in two-dimensional (2D)
nanodevices~\cite{wang12, Mak, Radisavljevic}, owing to the structural stability and lack of dangling bonds~\cite{Banerjee}. The ML-MoS$_2$ is a direct band gap semiconductor with a
band gap of $1.9$ eV~\cite{Mak}, and can be easily synthesized by using scotch tape or lithium-based intercalation~\cite{Mak, Radisavljevic, Banerjee, wang}. The mobility of the ML-MoS$_2$ can be at least $217$ cm$^2$V$^{-1}$s$^{-1}$ at room temperature using hafnium oxide as a gate dielectric, and the monolayer transistor shows the room temperature current on/off
ratios of $10^8$ and ultra low standby power dissipation~\cite{Mak}. These properties render Ml-MoS$_2$ as a promising candidate for a wide range of applications, including photoluminescence (PL) at visible wavelengths~\cite{photo}, and photodetectors~\cite{photodete}. The experimental achievements triggered the theoretical interests in the physical and
chemical properties of the ML-MoS$_2$ nanostructures to reveal the origins of the observed electrical, optical, mechanical, and magnetic properties,
and guide the design of MoS$_2$-based devices.

Having defined the valleytronics of graphene, many physical phenomena, originated from the spin of the electron, have been extended to be used for the valley index. One is the internal magnetic moments of spin which couples to an external magnetic field through well-known Zeeman interaction. In a system where the inversion symmetry is broken, the valley degree of freedom can be distinguished. There is a valley dependence orbital magnetic moment which can result in a Zeeman-like interaction for the valley index. Gapped graphene is one of the main representatives of materials in which the valley index couples to the perpendicular magnetic field as a real spin~\cite{ando}. However, due to the small value of the gap, this effect has not been yet observed experimentally. Transition metal dichalcogenides (TMDCs), on the other hand, provide a more applicable paradigm for the valley Zeeman (VZ) effect. The VZ in TMDCs has been recently observed~\cite{MacNill14,Aivizian14,Srivastava14,Li14} and studied theoretically~\cite{chu}. Those measurements were based on the shift of photoluminescence peak energies as a function of the magnetic field interpreted as a Zeeman splitting due to the valley-depended magnetic moments. In order to explore the VZ we do need to perceive all physical characteristics of the system. Actually, the energy band structure which can be calculated via \textit{ab-initio} methods, contains some information and besides, the Berry curvature and orbital magnetic moment of the Bloch states, are two main quantities which provide extra information to the band structure~\cite{Fuchs10,Niu96,Niu08}.

A peculiar property of the ML-MoS$_2$ is its spin-valley coupled electronic structure which is due to the strong spin-orbit coupling and it induces a spin-orbit splitting in the valence band~\cite{xiao12}.  Furthermore, many physical properties of TMDCs can be described by using a two-band model which is indeed a projected model from a higher dimension Hamiltonian. Since the projection is an approximation and it is not a perfect unitary transformation, the two-band Hamiltonian may not provide a full description of the low-energy excitations of the system especially when the system is addressed by a perpendicular magnetic field. Basically, some physics related to the multi-band structure such as Berry curvature and orbital magnetic moment properties might be ignored along the projection process. In this work, we would like to address these issues and explore their physical sources in the ML-MoS$_2$ structure.

An effective model based on a Dirac-like Hamiltonian has been introduced by Xiao et al.~\cite{xiao12} to explore ML-MoS$_2$ electronic properties. Very recently, it has been shown, based on the tight-binding~\cite{Rostami13,Liu13} and ${\bf k}\cdot{\bf p} $ method~\cite{Kormanyos13}, that a model going beyond the Dirac-like Hamiltonian (including effective mass asymmetry, trigonal warping, and a quadratic momentum dependent term) is very important. Each term in the Hamiltonian can be as a source of many physical consequences. For example, due to the spin-orbit coupling ($\lambda$) and the diagonal quadratic term ($\alpha$), the two-band model reveals a particle-hole asymmetry and also the diagonal quadratic term of $\beta$ gives a contribution to the Chern number at each valley~\cite{Rostami14}. A nanoribbon MoS$_2$ in the presence of the perpendicular magnetic field reveals the Landau level band structure with a VZ term~\cite{Rostami13}. We attempt to clarify the VZ concept based on symmetry arguments, semiclassical (orbital magnetic moment) and quantum mechanical (Landau levels) calculations. In other words, we emphasize that a particle-hole asymmetry originating from the orbital magnetic moment occurs in the presence of the perpendicular magnetic field and thus we express the physical reasons of the asymmetry observed in the experiments~\cite{MacNill14,Aivizian14,Srivastava14,Li14}.

In this paper, we further study the electronic structure and two-terminal electronic transport of a zigzag ML-MoS$_2$ in the presence of the perpendicular magnetic field. Our calculations are based on the multi-orbital tight-binding approach~\cite{Cappelluti13} which describes the electronic properties of the monolayer MoS$_2$ based on all \emph{d} and \emph{p} relevant orbitals of both the Mo and S atoms, respectively. We calculate the conductance of a clean and disordered systems in the presence of the perpendicular magnetic field by using a non-equilibrium recursive Green's function method~\cite{Datta}.

According to the spin-orbit coupling and the valley degeneracy breaking, a spin-valley polarization (SVP) is expected in the electronic structure of the bulk system and particularly in the hole doped case. Most remarkably, in the zigzag ribbon case, there are some metallic edge states which spoil the SVP in a clean system. However, our numerical results in the two-terminal conductance show a spin-valley polarized mode made by the quantum Hall and finite size edge states in the presence of on-site disorder.

The paper is organized as follows. In Sec.~II, we introduce the
formalism that will be used for calculating the electronic
structure, orbital magnetic moment, two terminal conductance and the valley polarization
quantity from the recursive Green's function approach. In Sec.~III,
we present our analytic and numeric results for the dispersion
relation in the presence of the magnetic field. Section~IV contains
a brief summary of our main results.

\section{theory and method}
\subsection{tight-binding model}
The tight-binding Hamiltonian is a common and a powerful technique to explore the transport properties. The model provides a reasonable description of the bulk properties of the ML-MoS$_2$ including direct band gap~\cite{Cappelluti13}. We carry out our calculations based on the following real space model Hamiltonian
\begin{align}
H&=\sum_{i,\mu}{\epsilon^a_{i,\mu}a^\dagger_{i,\mu}a_{i,\mu}+\epsilon^b_{i,\mu}({b^t}^\dagger_{i,\mu}{b^t}_{i,\mu}+{b^b}^\dagger_{i,\mu}{b^b}_{i,\mu})}
\nonumber\\&+\sum_{i,\mu}{[t^{\perp}_{i,\mu}{b^t}^\dagger_{i,\mu}{b^b}_{i,\mu}+ H.C.]}
\nonumber\\&+\sum_{\langle ij\rangle,\mu\nu}{[t^{ab}_{ij,\mu\nu} a^\dagger_{i,\mu}(b^t_{j,\nu}+b^b_{j,\nu})+H.C.]}
\nonumber\\&+\sum_{\langle\langle ij \rangle\rangle,\mu\nu}{[t^{aa}_{ij,\mu\nu} a^\dagger_{i,\mu}a_{j,\nu}+H.C.]}
\nonumber\\&+\sum_{\langle\langle ij\rangle\rangle,\mu\nu}{[t^{bb}_{ij,\mu\nu}({b^t}^\dagger_{i,\mu}{b^t}_{j,\nu}+{b^b}^\dagger_{i,\mu}{b^b}_{j,\nu})+H.C.]}\nonumber\\
\end{align}
where $\epsilon^a$ and $\epsilon^b$ indicate on-site energies for Mo and S atoms and $t^{ab},t^{aa}$ and $t^{bb}$ show the hopping
matrixes corresponding to Mo-S, Mo-Mo, and in-plane S-S hopping process, respectively. $t^{\perp}$ denotes the hoping integral between two sulfur layers, $i,j$ and $\mu,\nu$ stand for
lattice site and atomic orbital indices, respectively. Note that, the Hamiltonian is constructed by \emph{d} and \emph{p} orbitals of the Mo and S atoms which are listed as follows
\begin{align}
&d-\text{basis (Mo atoms)}:~~d_{z^2},~d_{x^2-y^2},~d_{xy},~d_{xz},~d_{yz}\nonumber\\
&p-\text{basis (S atoms)}:~~p_{x,t},~p_{y,t},~p_{z,t},~p_{x,b},~p_{y,b},~p_{z,b}
\end{align}
where the \emph{t} or \emph{b} subindex indicates the top or bottom sulfur plane, respectively. A unitary transformation is used to reduce the dimensionality of the Hamiltonian and thus relevant orbitals are only considered. The unitary matrix is given by
\begin{align}
U=\frac{1}{\sqrt{2}}\begin{pmatrix}I&&u\\I&&-u\end{pmatrix}
\end{align}
where $I$ is a three-dimensional identity matrix and $u={\rm diag}[1,1,-1]$. Implementing the unitary matrix on the $p$-basis of the sulfur atoms, results in two decoupled bases with a symmetric (even)
and an anti-symmetric (odd) combination of the \emph{p}-orbitals of two sulfur layers with respect to the horizontal reflection symmetry. These even and odd spaces read as
\begin{align}
&Even: \frac{1}{\sqrt{2}}(p_{x,t}+p_{x,b}),~\frac{1}{\sqrt{2}}(p_{y,t}+p_{y,b}),~\frac{1}{\sqrt{2}}(p_{z,t}-p_{z,b})\nonumber\\
&Odd:  \frac{1}{\sqrt{2}}(p_{x,t}-p_{x,b}),~\frac{1}{\sqrt{2}}(p_{y,t}-p_{y,b}),~\frac{1}{\sqrt{2}}(p_{z,t}+p_{z,b})\nonumber\\
\end{align}
The transformation gives rise to an opportunity to suppress direct coupling between two sulfur layers.
Based on the Hamiltonian in the main orbital space, two sulfur layers are directly coupled due to the vertical hopping as
\begin{align}
{\cal H}=\begin{pmatrix}h&&t^\perp\\t^\perp&&h\end{pmatrix}
\end{align}
where $h=\epsilon^b$ which indicates the on-site term of the tight-binding Hamiltonian corresponding to the $p$-orbitals of the Sulfur atoms on both top and bottom layers. Using $u t^\perp=t^\perp u$, and $ u\epsilon^b=\epsilon^b u$ one can show that in the new space we have
\begin{align}\label{hp}
{\cal H'}=U {\cal H} U^\dagger=\begin{pmatrix}h+ut^\perp&&0\\0&&h-ut^\perp\end{pmatrix}
\end{align}
where the first ($\tilde\epsilon^b=\epsilon^b+ut^\perp$) and second diagonal block belong to the even and odd symmetric subspaces~\cite{Cappelluti13}, respectively. Therefore, the six-band real space Hamiltonian can be written in the even symmetric subspace which contains even subspace of p-orbital and even subspace of d-orbital (i.e. $d_{z^2},~d_{x^2-y^2},~d_{xy}$). Besides, in the presence of the perpendicular magnetic field, the six-band Hamiltonian reads as
\begin{align}\label{r-space}
H&=\sum_{i,\mu}{\epsilon^a_{i,\mu}a^\dagger_{i,\mu}a_{i,\mu}+\tilde{\epsilon}^b_{i,\mu}b^\dagger_{i,\mu}b_{i,\mu}}
\nonumber\\&+\sum_{\langle ij\rangle,\mu\nu}{[e^{i\phi_{ij}}t^{ab}_{ij,\mu\nu} a^\dagger_{i,\mu}b_{j,\nu}+ H.C.]}
\nonumber\\&+\sum_{\langle\langle ij\rangle\rangle,\mu\nu}{[e^{i\phi_{ij}}t^{aa}_{ij,\mu\nu} a^\dagger_{i,\mu}a_{j,\nu}+H.C.]}
\nonumber\\&+\sum_{\langle\langle ij\rangle\rangle,\mu\nu}{[e^{i\phi_{ij}}t^{bb}_{ij,\mu\nu}b^\dagger_{i,\mu}b_{j,\nu}+H.C.]}\nonumber\\
\end{align}
Using Eq.~(\ref{hp}), together with the crystal fields of the system~\cite{Cappelluti13}, and also spin-orbit couplings for the valence and conduction bands in atomic limit, i.e. ${\bf L}\cdot{\bf S}$, the on-site energy matrices are given by
\begin{align}
\epsilon^a_{i,\mu}&=\begin{pmatrix}\Delta_0&&0&&0\\0&&\Delta_2&&-i\lambda_M\hat{s}_z\\0&&i\lambda_M\hat{s}_z&&\Delta_2\end{pmatrix}\nonumber\\ \nonumber\\
\tilde{\epsilon}^b_{i,\mu}&=\begin{pmatrix}\Delta_p+t^\perp_{xx}&&-i\frac{\lambda_X}{2}\hat{s}_z&&0\\i\frac{\lambda_X}{2}\hat{s}_z&&\Delta_p+t^\perp_{yy}&&0\\0&&0&&\Delta_z-t^\perp_{zz}\end{pmatrix}
\end{align}
where $\lambda_M=0.075$eV and $\lambda_X=0.052$eV stand for the spin-orbit coupling originating from the Mo (metal) and S (chalcogen) atoms, respectively~\cite{Roldan14}. Notice that $s=\pm$ indicates the $z-$component of the spin degree of freedom. Moreover, we have added an external perpendicular magnetic field to the system using Peierls phase factor, $\phi_{ij}=\frac{e}{\hbar}\int_i^j{\vec{A}\cdot\vec{dr}}$ to carry out the orbital effect of the perpendicular magnetic field. Interlayer hopping between the Sulfur planes is given as $t^\perp={\rm diag}[V_{pp\pi},V_{pp\pi},V_{pp\sigma}]$ based on the Slater-Koster table~\cite{sk}.
The numerical values of the tight-binding parameters are $\Delta_0=-1.096$, $\Delta_2=-1.512$, $\Delta_p=-3.560$, $\Delta_z=-6.886$,  $V_{dd\sigma}=-0.895$, $V_{dd\pi}=0.252$, $V_{dd\delta}=0.228$ ,$V_{pp\sigma}=1.225$,$V_{pp\pi}=-0.467$, $V_{pd\sigma}=3.688$, and $V_{pd\pi}=-1.241$ in eV units. These parameters will be presented elsewhere [~\onlinecite{Rostami142}]. We might express that this Hamiltonian provides a very good energy band structure in according to the comparison with those results obtained within the density functional theory simulations~\cite{Roldan14}.

\subsection{Orbital magnetic moments}

In many semiconductor systems, such as GaAs bulk, the circular polarization of luminescence from circularly polarized excitation originates from electron or hole spin polarization~\cite{parsons}. However in ML-MoS$_2$, the optical selection rule originates from the orbital magnetic moments at each, $K$ or $K'$, valley independent of electron or hole spin~\cite{yang}.

In a periodic lattice, the eigenfunctions of the Schr\"{o}dinger equation are Bloch states $u_{n, k} $, where $n$ and $k$ indicate the band index and crystal momentum, respectively. In semiclassical method, it is common to use a wave packet picture of electrons~\cite{Fuchs10,Niu96,Niu08}. The wave packet, $|W\rangle$, can be easily constructed by the linear superposition of the Bloch states. Due to the self rotation of the wave packet around its own center of mass, the magnetic moment (or the angular orbital momentum ${\bf L}$) defined as ${\bf M}=-\frac{e}{2m_0}{\bf L}=-\frac{e}{2m}\langle W|(\hat{r}-r_c)\times \hat{p}|W\rangle$ along the $z-$direction where $m_0$ is the free electron mass and $\hat{p}$ is the canonical momentum operator and moreover the wave packet is also centered at $r_c$ in the position space. The orbital magnetic moment of Bloch electrons has a contribution from inter cellular current circulation governed by symmetry properties. After straight forward calculations~\cite{Fuchs10,Niu96,Niu08}, the orbital magnetic moment is written as
\begin{align}
{\bf M}_n(k)=i\frac{e}{\hbar}\sum_{m\neq n}\langle \nabla_{k} u_{n k}|\times[H(k)-\epsilon_{n k}]|\nabla_{k} u_{n k}\rangle
\end{align}
This relation can be written in a more practical expression as
\begin{align}\label{eq:OMM}
{\bf M}_n(k)=-\hat{z}\frac{e}{\hbar}\sum_{m\neq n} \frac{{\rm Im}[\langle u_{n k}|\partial_{k_x}H(k)|u_{m k}\rangle \langle u_{m k}|\partial_{k_y}H(k)|u_{n k}\rangle]}{\epsilon_{n k}-\epsilon_{m k}}
\end{align}
Up to linear order in the magnetic field and in semiclassical limit, the energy dispersion in an external magnetic field modifies as
\begin{align}
E_{nk}=\epsilon_{nk}-{\bf M}_n(k)\cdot {\bf B}
\end{align}
where $\epsilon_{nk}$ is the band dispersion of the system without magnetic field. It is worth to mention that the inversion and time reversal symmetries play vital roles in the nontrivial Berry curvature and the orbital magnetic moment. According to the time reversal symmetry, ${\bf M}(k)=-{\bf M}(-k)$ while the presence of the inversion system results ${\bf M}(k)={\bf M}(-k)$. Consequently, the orbital magnetic moment vanishes by governing both symmetries. Most importantly, the magnetic moment is non-zero in ML-MoS$_2$ since the inversion symmetry is broken. Similar behavior is expected for the Berry curvature as well. In order to calculate the orbital magnetic moment, based on the six-band tight-binding model, we carry out a Fourier transformation along the $x$ and $y$ directions to find the six-band Hamiltonian in the $k$-space. Moreover, the orbital magnetic moment can be also found through the corresponding two-band model around the $K-$point. The two-band model can be extracted by using L\"{o}wding partitioning method from the six-band Hamiltonian. The two-band Hamiltonian of the monolayer MoS$_2$, after ignoring the trigonal warping and the momentum dependence of the spin-orbit coupling, is given by
\begin{eqnarray}\label{hk}
H&=&\frac{\Delta_0+\lambda_0\tau s}{2}+\frac{\Delta+\lambda\tau s}{2}\sigma_z\nonumber\\&+&t_0 a_0 {\bf q}\cdot{\bm \sigma}_\tau+\frac{\hbar^2|{\bf q}|^2}{4m_0}(\alpha+\beta\sigma_z)
\end{eqnarray}
where $s=\pm$ and $\tau=\pm$ indicate spin and valley, respectively  ${\bm \sigma}_{\tau}=(\tau\sigma_x,\sigma_y)$ are Pauli matrices and  ${\bf q}=(q_x, q_y)$ is momentum. The numerical values of the two-band model parameters are given by $\Delta_0=-0.11$eV, $\Delta=1.82$eV, $\lambda_0=70$meV, $\lambda=-80$meV, $t_0=2.33$eV, $\alpha=-0.01$, and $\beta=-1.54$. The $z-$component of the orbital magnetic moment of the conduction and valence bands in the two-band model Hamiltonian are given by
\begin{align}\label{Eq:m2}
M^s_c(k)=M^s_v(k)=-\tau\frac{ e}{\hbar}\frac{t_0^2 a_0^2(\Delta -2 b \beta a_0^2 k^2+\lambda s)}{(\Delta +2 b \beta a_0^2  k^2+\lambda s )^2+4 t_0^2 a_0^2 k^2 }
\end{align}
where $b=\hbar^2/4m_0a_0^2\approx 0.572$. Moreover, at two valleys ($k=0$) the contribution from $\beta$ is eliminated and one can find
\begin{align}
M^s_c(k=0)=M^s_v(k=0)=-\tau\frac{e}{\hbar}\frac{ t_0^2a_0^2}{\Delta+\lambda s}
\end{align}
Note that for the low-energy model parameters, we have $\hbar M^{\uparrow}(k=0)/(e a_0^2)\approx -3.14 \tau $ eV and $\hbar M^{ \downarrow}(k=0)/(e a_0^2)\approx -2.87 \tau $ eV.

It should be noticed that the opposite sign of the orbital magnetic moments at two valleys, which originates from the time reversal symmetry, leads to the VZ effect when the system is imposed by an external perpendicular magnetic field. Moreover, the low-energy Hamiltonian exhibits the same value of the {semiclassical} magnetic moment at both the valence and conduction bands while the recent experimental studies showed a different value for the magnetic moment at two bands. In the numerical section, we will discuss this discrepancy more carefully.

Although the magnitude of the valley splitting in each band has not been measured experimentally, the mismatch was measured in four different experiments. The photoluminescence intensity of a monolayer transition metal dichalcogenide has been measured in the presence of the external perpendicular magnetite field using circular polarized light as the excitation light. The shift value of the peak of the luminescence spectrum of MoSe$_2$ \cite{MacNill14,Li14} and WSe$_2$ \cite{Aivizian14,Srivastava14} are about $2-5$ meV for left- and right-handed polarizations and for both neutral and charged exciton.

The linear dependence of the valley splitting demonstrates a Zeeman-like effect of the valley index. According to the circular dichroism effect in these materials, the right- (left-) handed light couples just to the $K$ ($K'$) valley. In the magnetic field the energy gap between electron and hole states differs in two valleys, whereas $E_{\rm CBM}-E_{\rm VBM}=\Delta+\lambda+\tau(g_v^{con}-g_v^{val})\hbar\omega_c/2$ and the difference provides an opportunity to the valley Zeeman effect to be measured experimentally. Therefore, due to the circular dichroism effect, the left- and right-handed emitted light have two different frequencies (i.e. corresponding energy gap) leading to a splitting in the peak of the PL spectrum for two polarizations.

Being aware of the discrepancy of the two-band model in the magnetic field and in order to capture the correct value of the orbital magnetic moment of the system, we add a mismatch, $\kappa_v$ between the semiclassical orbital magnetic moments of the six- and two-band models at the $K$-point to the low-energy two-band Hamiltonian when there is a perpendicular magnetic field. Consequently, in the presence of the magnetic field, the low-energy Hamiltonian, Eq. (12), is modified as
\begin{eqnarray}
{\cal H}_{\tau s}&=&\frac{\Delta_0+\lambda_0\tau s}{2}+\frac{\Delta+\lambda\tau s}{2}\sigma_z+v{\bm \pi}\cdot{\bm \sigma}_\tau
+\frac{|{\bm \pi}|^2}{4m_0}(\alpha+\beta\sigma_z)\nonumber\\&-&\frac{1}{2}\tau\kappa_v\hbar\omega_c-\frac{1}{2}sg_s\hbar\omega_c\nonumber \\
\end{eqnarray}
where ${\bm \pi}={\bf p}+e{\bf A}$ and $g_s\approx2$ is the Zeeman coupling for the real spin and the mismatch between the Zeeman coupling of both the bands is
\begin{eqnarray}
\kappa_v&=&\frac{1eV}{\hbar^2/(4m_0a_0^2)}\begin{pmatrix}m_c-m_2&&0\\0&&m_v-m_2\end{pmatrix}\nonumber\\&\approx&
\begin{pmatrix}-0.62&&0\\0&&-1.50\end{pmatrix}
\end{eqnarray}
where $m_2$ (in unit of $e^2 V a_0^2/\hbar$) is the magnetic moment calculated by the two-band model while $m_c$ and  $m_v$ are the magnetic moment obtained within the six-band tight-binding model in the conduction and valence bands, respectively. The numerical values of $\kappa_v$ (which is about $\hbar\omega_c=\hbar (eB/2m_0)$) are obtained by using the semiclassical results of the orbital magnetic moments presented in Fig. 3 at the $K$ point and by averaging over spins. We also define $\kappa^{con}_v=-0.62$ and $\kappa^{val}_v=-1.5$.

\subsection{Conductance and spin-valley polarization}

Using the Fourier transformation along the ribbon, the energy dispersion can be found as $H_k=H_{00}+H_{01}e^{i ka}+H_{01}^\dagger e^{-i ka}$
where $H_{00}$ and $H_{01}$ are the intra and inter principal cell Hamiltonian,
respectively~\cite{Rostami132}. Note that $a=\sqrt{3}a_0=0.316$nm stands for the Mo-Mo or in-plane the S-S bond length with $a_0$ as the in-plane projection the Mo-S bond length.
To calculate the conductance, we use the non-equilibrium Green's function
method in which the retarded Green's function is defined as
$G^r_s=(E-H_s-\Sigma_s+i0^+)^{-1}$ by employing the recursive Green's
function method~\cite{negf02}. Note that $s=\uparrow \text{or} \downarrow$ for the spin degree of freedom.
In the noninteracting Hamiltonian,
the self-energy ($\Sigma_s=\Sigma^L_s+\Sigma^R_s$) originates only from
the connection of the system to leads and it can be calculated
by the method that has been developed by Lopez et al~\cite{lopez}.
Using the Landauer formula, the zero temperature conductance for each spin component is given as $G_{\uparrow(\downarrow)}=\frac{e^2}{h}T_{\uparrow(\downarrow)}$ where
\begin{eqnarray}\label{cond}
T_{s}={\rm Tr}[\Gamma^L_s G^r_s \Gamma^R_s G_s^{r\dagger}]
\end{eqnarray}
and $\Gamma^{L,R}_s=-2\Im[\Sigma^{L,R}_s]$ are line width functions. Because of the collinear spin structure, the conductance of each spin component
can be calculated separately. Consequently, in principal, a spin polarization quantity can be defined as $P=(G_{\uparrow}-G_{\downarrow})/(G_{\uparrow}+G_{\downarrow})$.
\begin{figure}
\includegraphics[width=1\linewidth]{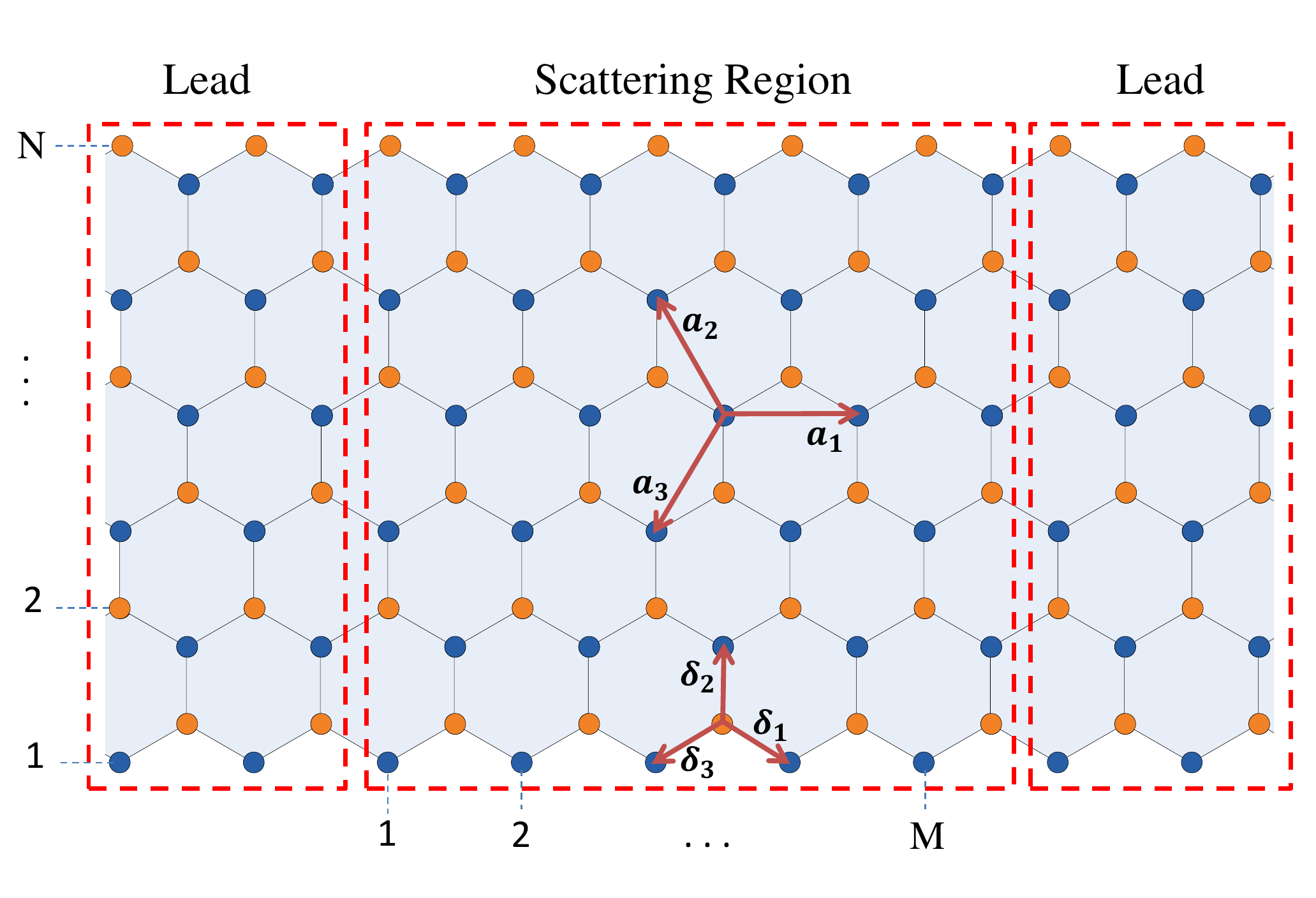}
\caption{(Color online) A top view schematic of a monolayer MoS$_2$ lattice structure in a two-terminal setup. Blue (orange) circles indicate the \emph{Mo} (\emph{S}) atoms. The nearest neighbor ($\delta_i$) and the next nearest neighbor ($a_i$) vector are shown in the figure. Ribbon width and scattering region length are $W/a_0=3N/2-1, L/a_0=\sqrt{3}M$, respectively.}
\label{scheme}
\end{figure}

\section{Results and discussion}

In this section, we present our main results in the orbital magnetic moment, Landau levels spectrum and spin-valley polarized transport in monolayer MoS$_2$ in the presence of the perpendicular magnetic field. We present our extensive numerical results of the electronic structure by exploring the structure of the Landau levels in the quantum Hall regime and the spin-valley resolved transport properties of the zigzag MoS$_2$ nanoribbon. We calculate the conductance in both unipolar electron and hole doped cases and we explore the spin-valley-resolved electronic transport in both clean and disordered systems.

\subsection{Valley Zeeman and Landau levels}

\begin{figure}
\centering
\includegraphics[width=0.95\linewidth]{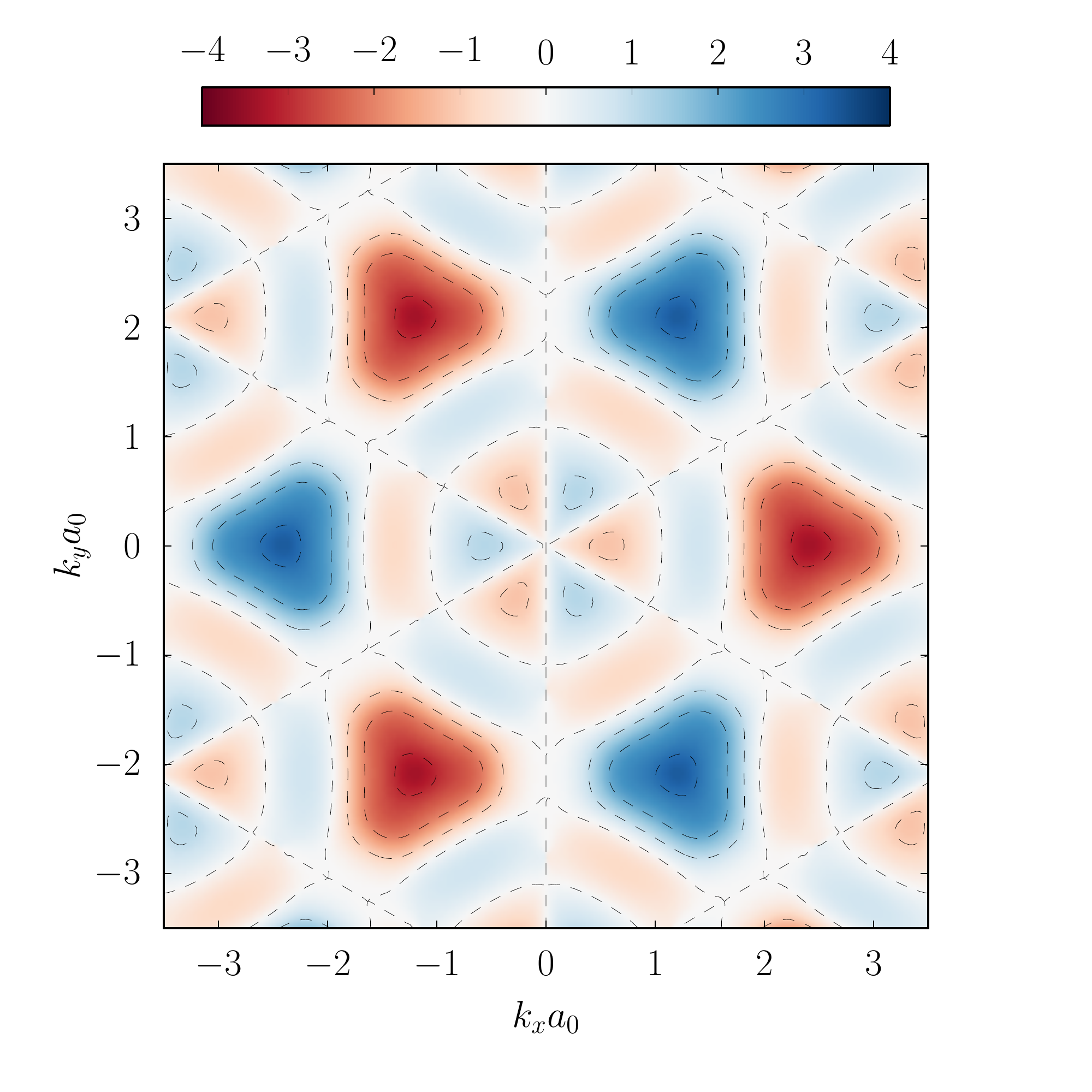}
\includegraphics[width=0.95\linewidth]{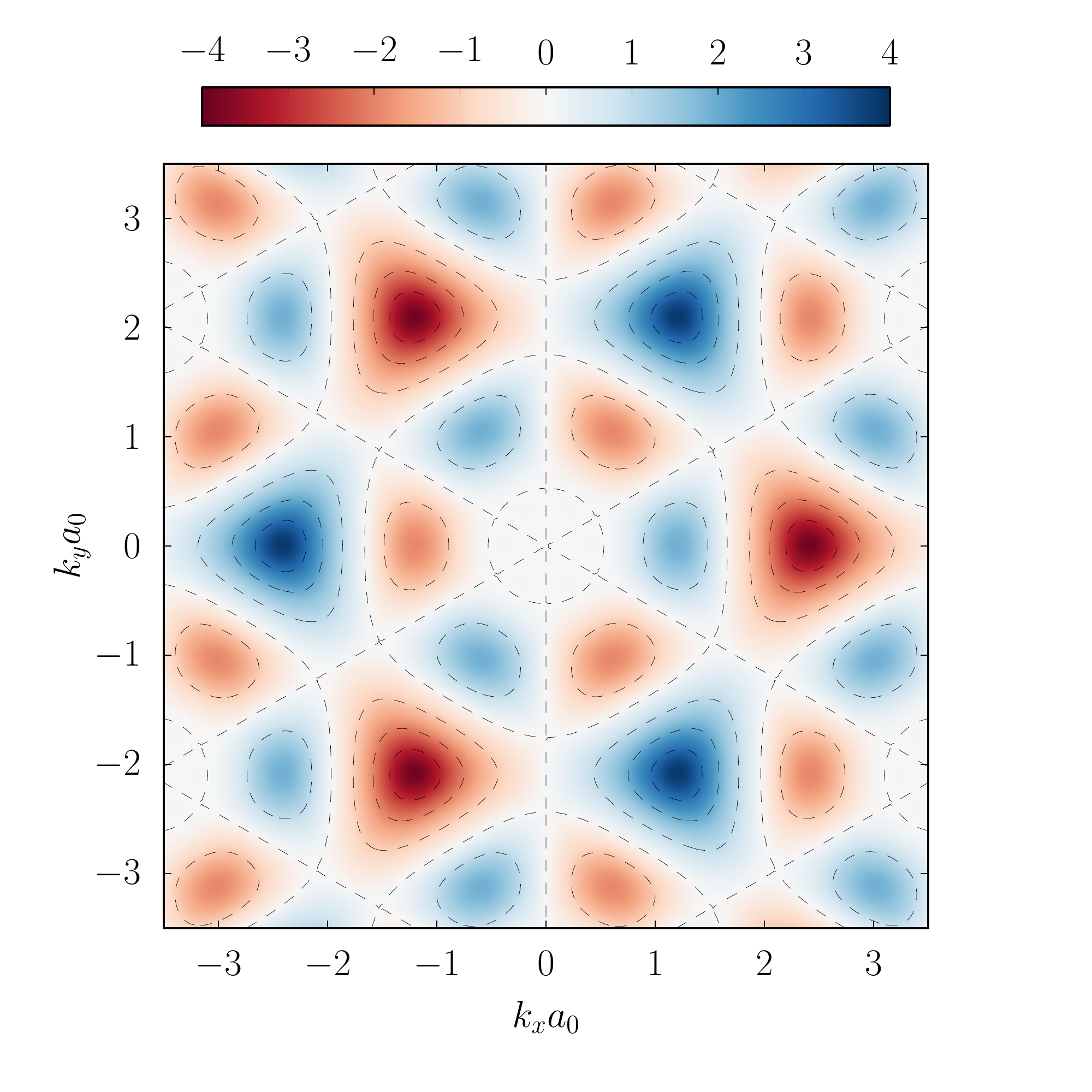}
\caption{(Color online) Contour plot of the orbital magnetic moment as function of the momenta along the $x$ axis at the conduction (top panel ) band and the valence (below panel) band. $M$ is in unit of $e^2 V a_0^2/\hbar$ and the spin orbit coupling is neglected in this figure.}
\label{fig:omm1}
\end{figure}

\begin{figure}
\centering
\includegraphics[width=0.95\linewidth]{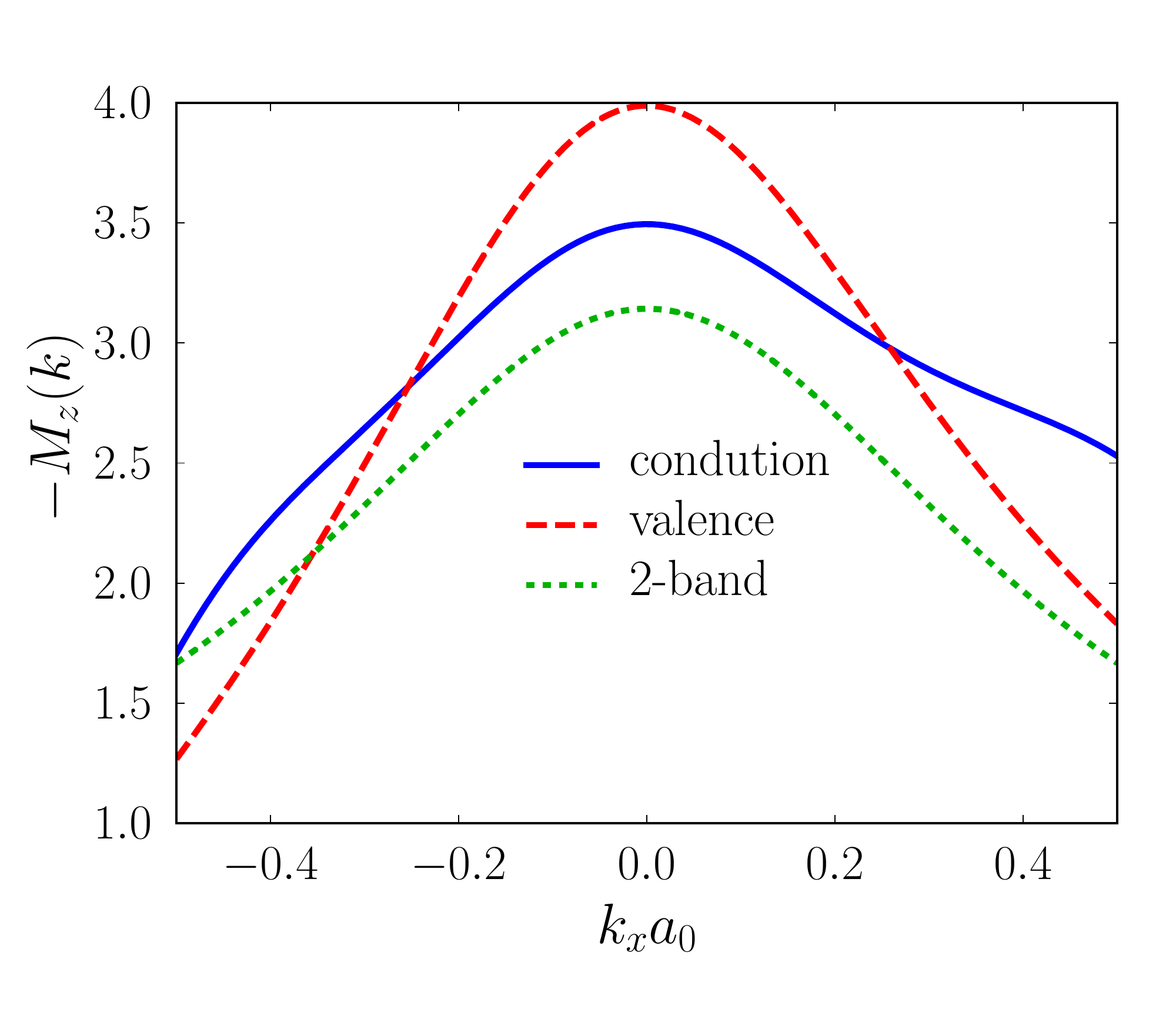}
\includegraphics[width=0.95\linewidth]{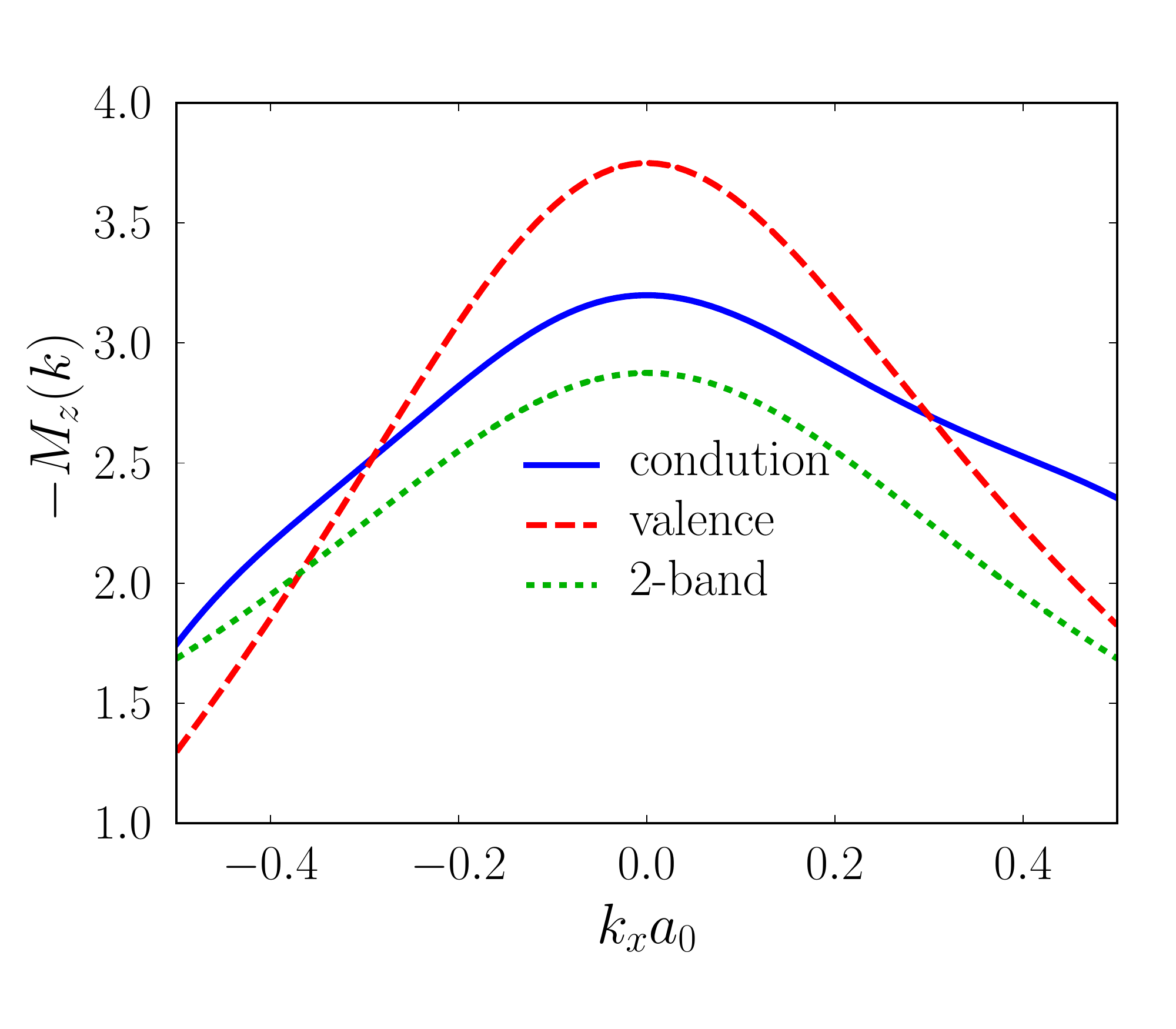}
\caption{(Color online) Orbital magnetic moment as a function of the momentum along the $x-$ axis for both the spin up and down components calculated by the six-band and the two-band models. Up (bellow) panel corresponds to the spin up (down) component and $M$ is in unit of $e^2 V a_0^2/\hbar$.}
\label{fig:omm2}
\end{figure}

Before calculating the conductance of the system, we first discuss the VZ effect induced by the perpendicular magnetic field in both semiclassical and quantum aspects. First of all, the orbital magnetic moment corresponding to the conduction and valence bands are calculated in the whole Brilloun zone (BZ) using the six-band tight-binding model, specially using Eqs.~(7-8) and (10), and results are shown in the counter plots in Fig.~\ref{fig:omm1}. It is obvious that the orbital magnetic moment changes sign in the two valleys owing to the time reversal symmetry. Indeed, the states near the corners of the BZ contribute mainly to the orbital magnetic moment. Moreover, a comparison between the semiclassical orbital magnetic moment calculated within the two-band, using Eq. (\ref{Eq:m2}), and the six-band models as a function of the momentum along $x$ axis are shown in Fig.~\ref{fig:omm2} for both spin components. As seen in the figure, a remarkable difference between the value of the orbital magnetic moment in the valence and conduction bands is obtained by the six-band model Hamiltonian. However, in the two-band model, the semiclassical magnetic moment is the same in both the valence and conduction bands (see Eq.~\ref{Eq:m2}) even in the presence of the particle-hole asymmetry terms such as the spin-orbit coupling and effective mass asymmetry. Most remarkably, the mismatch between the orbital magnetic moment of two bands calculated within the six-band model plays an important role in interpreting the VZ experimental measurements.

The difference between the two- and six-band models can be classified in two intraband and interband categories. The intraband reason is related to the orbital character of the bands. Using the Slater-Koster table for constructing the tight-binding model, provides a platform for taking into account the nature of the relevant atomic orbitals such as $p$ and $d$ types and also considering the neighboring lattice symmetry. However the orbital basis of the two-band model is substituted with the band basis and the orbital character can be mainly captured by $d$-type orbitals.

According to Eq.~(\ref{eq:OMM}), similar to the Berry curvature formula and the second order perturbation theory, the orbital magnetic moment of each band is affected by virtual transitions between bands corresponding to the inter-band sector~\cite{andor}. Due to the transition between neighboring energy bands, observing different value of the orbital magnetic moment of two different bands is awaited, however such virtual transition is definitely eliminated in the two-band case. Consequently, we would like to emphasize that one might be careful in using the L\"{o}wdin canonical projection from a multi-band to a two-band model, because some information regarding the orbital character and virtual transitions might be ignored.

The wave vector point group symmetry of a honeycomb lattice with broken inversion symmetry, like gapped graphene, is $C_{3h}$ point group~\cite{Inoue11,Ochoa13} near the $K$ and $K'$ points. The irreducible representations of the point group characterize energy eigenfunctions at the $K$ and $K'$ valleys. According to the character table, the phase winding at each $K$ and $K'$ is $C_3|c,\tau\rangle=\omega^{\tau}|c,\tau\rangle$ and $C_3|v,\tau\rangle=\omega^{-\tau}|v,\tau\rangle$ where $\omega=e^{i2\pi/3}$ due to three-fold rotational for the conduction and the valence bands. The relation means that the orbital angular momentum in the conduction band is $l_c=-\tau$ and similarly $l_v=\tau$ for the valence band. In a semiclassical picture, the angular momentum has been induced from the self-rotation of the electron wave packet around its center of mass. This kind of the orbital angular momentum, called Bloch phase shift, is well studied in the content of gapped graphene which can be explained by a single $p_z$-orbital tight-binding model. However, in any multi-orbital system, another distinct contribution to the orbital angular moment might be expected.

At high symmetric points where the Bloch states are invariant under a $g-$fold discrete rotation, an azimuthal selection rule $l_c+gN=l_v \pm1$ is expected for interband transitions. According to the {\it ab-initio} calculations near the $K (K')$ point, the conduction band minimum is mainly formed from the Mo $d_{z^2}$ orbitals with $l_z=0$ and the valence band is constructed by the Mo $d_{x^2-y^2}+id_{xy}$ ($d_{x^2-y^2}-id_{xy}$) orbital with $l_z=2$ ($l_z=-2$). Note that there are some contributions from $p_x$ and $p_y$ orbitals of the S atoms in both band edges. If the mixing from $p-$orbital is ignored, the total angular momentum will be $l_c\sim-\tau$ and $l_v\sim\tau+2\tau\sim3\tau$ including the Bloch phase shift and local orbital contribution of the conduction band. Moreover, owing to the selection rule allowed with discrete three-fold rotational symmetry, we can add a multiplicand of three to the orbital angular moment of one of the bands in order to satisfy $l_v-l_c=\pm1$ which is necessary in the dipole absorption limit~\cite{yao08}. In this case, we have $l_v\sim0$ and $l_c=-\tau$.

\begin{figure}
\includegraphics[width=0.91\linewidth]{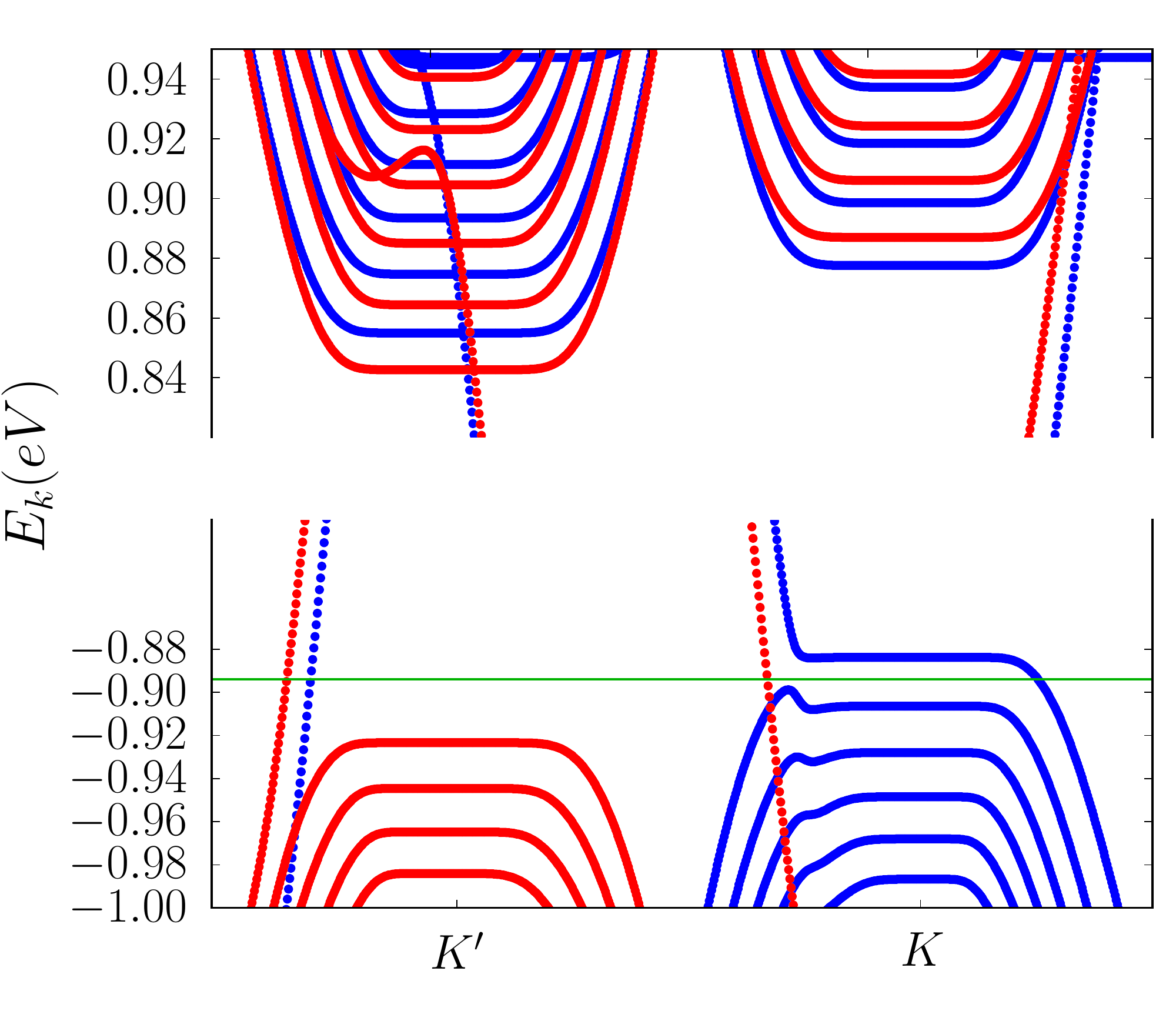}
\includegraphics[width=0.97\linewidth]{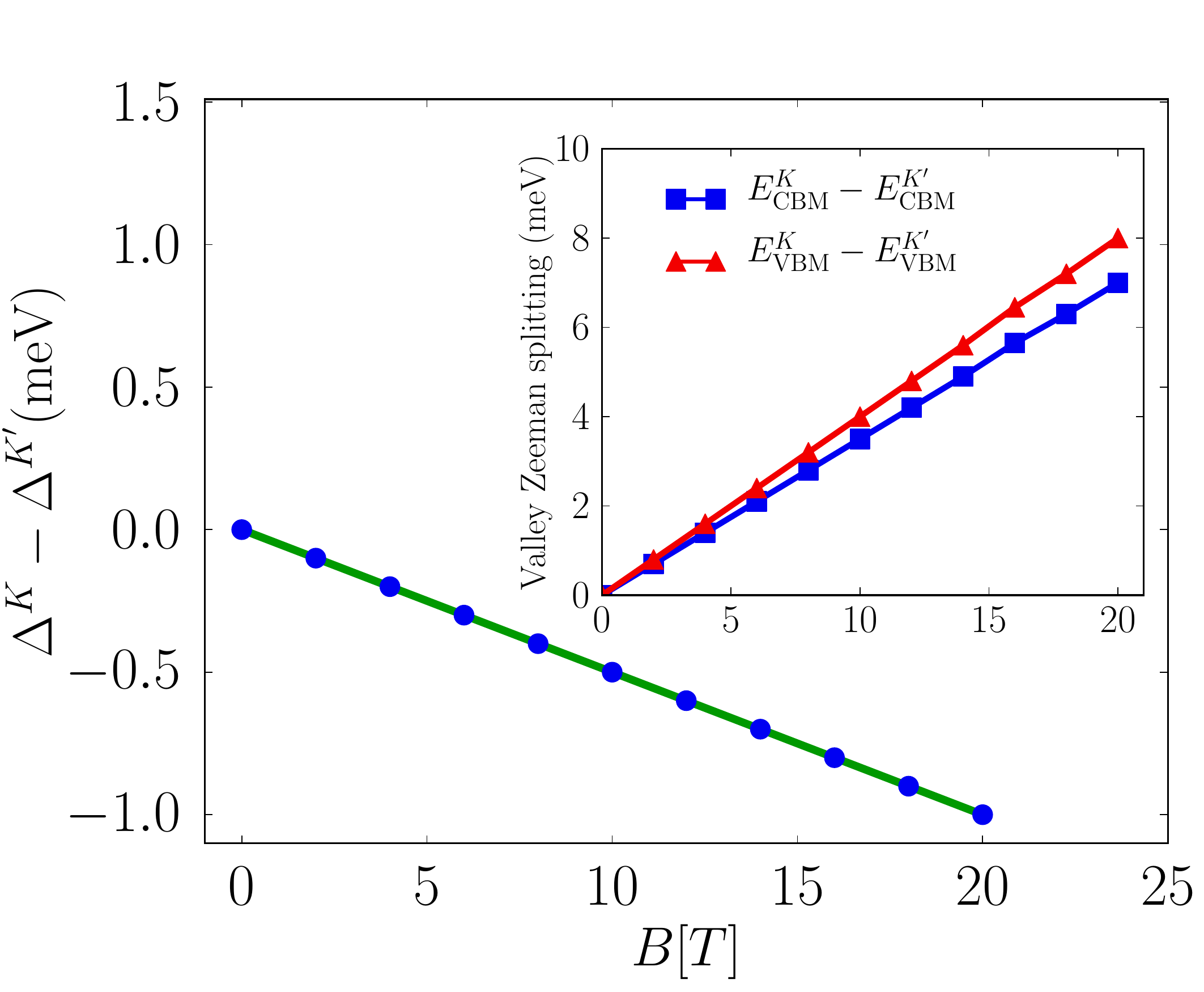}
\caption{(Color online) (top panel) Landau levels as a function of the momentum in unit of eV calculated by tight-binding approach on a zigzag ribbon where $B=100$T. (bottom panel) Valley Zeeman splitting in unit of meV as a function of the magnetic field in unit of tesla for both the conduction and valence bands. In the inset: the mismatch between the valley Zeeman effect of the conduction and valence bands which is the splitting in PL spectrum for right and left handed polarized light as a function of the magnetic field in unit of tesla. Note that blue (red) lines indicate spin up (down) states. We set $N=100$ as the ribbon width and the real Zeeman effect is not included in this figure.}
\label{fig:LL2}
\end{figure}

The Landau level spectrum is also calculated within the six-band model (see Fig.~\ref{fig:LL2}) of a zigzag ribbon ML-MoS$_2$ after applying a Peierls substitution in the tight-binding model. Thus, by using the Landau level spectrum resulted from full tight-binding calculation, we extract the valley Zeeman effect of the conduction and valence bands. The mismatch between the splitting in two bands, which is the shift of the PL spectrum of right- and left-handed light in the presence of the magnetic field, is shown in Fig.~\ref{fig:LL2} (bottom panel). This linear dependence of the magnetic field magnitude of the energy splitting approves the Zeeman-like coupling and is in good agreement with those results measured in experiments.

Having calculated the orbital magnetic moments in the six- and two-band models, we modified the two-band model Hamiltonian in the presence of the perpendicular magnetic field given by Eq. (15). After a straight forward calculation, the Landau level spectrum of the modified two-band Hamiltonian, Eq.~(15) reads as
\begin{align}\label{ELL}
\centering
E^\pm_{n\neq0,\tau s}&=\pm\sqrt{[\frac{\Delta+\lambda \tau s}{2}+ \hbar \omega_c (\beta n-\frac{ \alpha\tau }{2})]^2+2(\frac{t_0 a_0}{l_B})^2n}\nonumber\\
&+\frac{\Delta_0+\lambda_0\tau s}{2}+\hbar \omega_c(\alpha n -\frac{\beta\tau}{2})-\frac{1}{2}\tau\kappa_v\hbar\omega_c\nonumber\\&-\frac{1}{2}sg_s\hbar\omega_c\nonumber\\
E^{-}_{n=0,K s}&=\frac{\Delta_0+\lambda_0 s}{2}-\frac{\Delta+\lambda s}{2}+\frac{\hbar\omega_c}{2}(\alpha-\beta)\nonumber\\&-\frac{1}{2}\kappa^{val}_v\hbar\omega_c-\frac{1}{2}sg_s\hbar\omega_c\nonumber\\
E^{+}_{n=0,K's}&=\frac{\Delta_0-\lambda_0s}{2}+\frac{\Delta-\lambda s}{2}+\frac{\hbar\omega_c}{2}(\alpha+\beta)\nonumber\\&+\frac{1}{2}\kappa^{con}_v\hbar\omega_c-\frac{1}{2}sg_s\hbar\omega_c
\end{align}
in the presence of a constant magnetic field ${\bf B}$. It must be noticed that for $n=0$ level, there is no solution of the eigenvalue problem in the conduction band at the $K$-point and similarly in the valence band at the $K'$-point. Having calculated the analytical expression of the Landau level from the two-band model, we could deduce a valley splitting the conduction band and adding the contribution from a real Zeeman interaction and multi-band correction. The valley splitting coupling in the conduction and valence bands can be defined as $g^{con}\hbar\omega_c=E^{+}_{1,K \uparrow}-E^{+}_{0,K'\downarrow}$ and $g^{val}\hbar\omega_c=E^{-}_{0,K \uparrow}-E^{-}_{1,K'\downarrow}$, respectively with the following explicit expressions
\begin{align}
\centering
g^{con(val)}\hbar\omega_c&=\sqrt{[\frac{\Delta+\lambda}{2}+ \hbar \omega_c (\beta \mp \frac{ \alpha}{2})]^2+2(\frac{t_0 a_0}{l_B})^2}
\nonumber\\&- \frac{\Delta+\lambda}{2}-\hbar \omega_c(\beta\mp\frac{\alpha}{2})-(\kappa^{con(val)}_v+g_s)\hbar\omega_c
\end{align}
where $-/+$ stands for the conduction/valence band.
This is important that $\alpha$ fhhas no effect on the semiclassical orbital magnetic moment while it is a source of the mismatch of the magnetic moment (i.e. valley splitting) in those bands from a quantum point of view. In other words, in the quantum picture, the two-band model could produce a mismatch between magnetic moments while this is not the case in the semiclassical picture.
It is worth to expand above relation up to leading order in a weak magnetic field as
\begin{align}\label{gvalley}
g^{con,val}&\approx \frac{4
   a_0^2 m_0 t_0^2}{\hbar^2 (\Delta +\lambda )}+\frac{2 a_0^2 m_0 t_0^2 \left(\frac{(\pm\alpha -2 \beta ) (\Delta +\lambda )}{m_0}-\frac{4 a_0^2 t_0^2}{\hbar^2}\right)}{l_B^2(\Delta +\lambda )^3}\nonumber\\&
-\kappa^{con,val}_v- g_s
\end{align}
Here, using the six-band tight-binding model, the relation for the splitting is given by
\begin{align}
g^{con}-g^{val}=\frac{4  a_0^2 e t_0^2}{\hbar (\Delta+\lambda )^2}\times \alpha\times B-(\kappa^{con}_v-\kappa^{val}_v)
\end{align}
It is clear that the effective mass asymmetry (i.e. $\alpha$) yields a quadratic dependence of the mismatch to the magnetic field which can compete with the diamagnetic shift of the exciton binding energies which is also quadratic in $B$ ~\cite{Bayer,Gippius,Walck}. However, that can not explain those PL experimental data while the correction from the multi-band and the multi-orbital nature of this material ($\kappa_v$) gives rise a linear shift of the PL spectrum of left- and right-handed light. Therefore, our low-energy model predicts $g^{con}-g^{val}\sim -0.88+\frac{7.22 a_0^2}{l_B^2}\alpha$. Based on the tight-binding model, Fig.~4 bottom panel, $g^{con}-g^{val}\sim -0.81$ indicating that the proposed Eq.~(\ref{gvalley}) is reasonably good by incorporating the semiclassical approach of the value $\kappa^{con}_v$ and $\kappa^{val}_v$.

\begin{figure}
\includegraphics[width=0.95\linewidth]{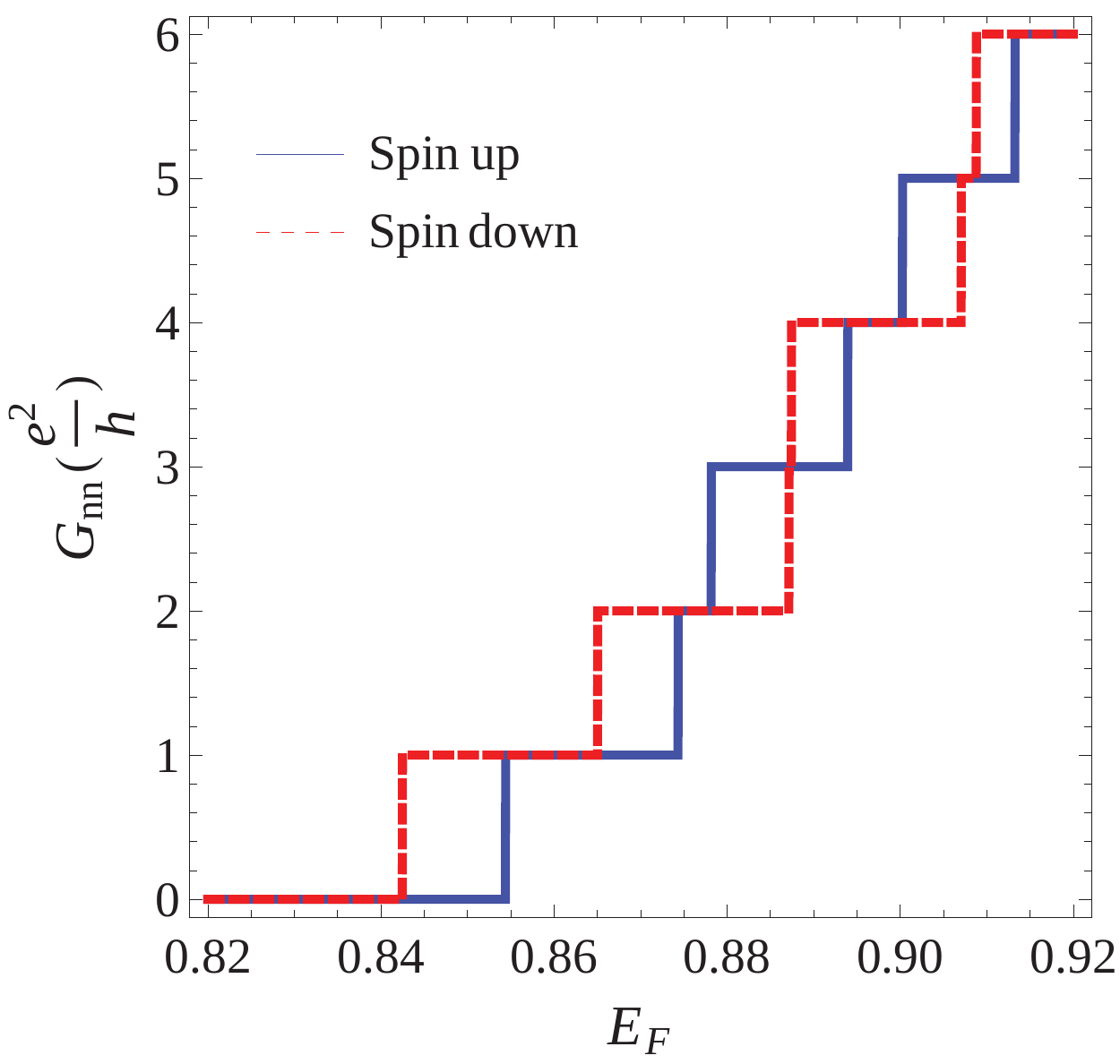}
\includegraphics[width=0.95\linewidth]{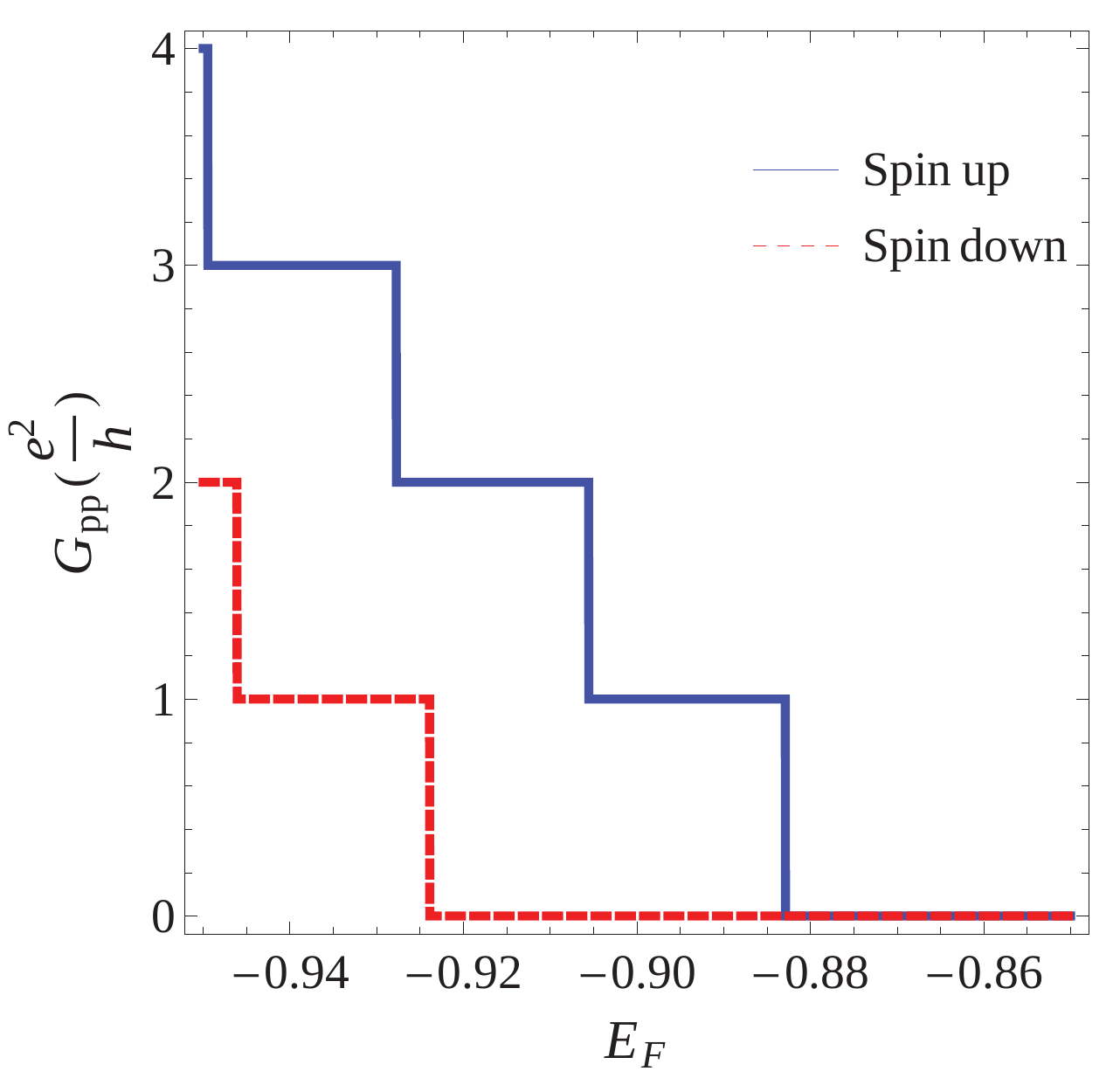}
\caption{(Color online) Unipolar conductance according to the Landau level spectrum of the low-energy model. The Zeeman interaction corresponding to the real spin is not taken into account in this figure.}
\label{fig:cond1}
\end{figure}

\subsection{Spin polarization: two-terminal transport}

The optical probing such as the PL approach can just measure the mismatch between the valley Zeeman effect of electron and hole states since measuring valley Zeeman splitting at each band requires a transition between two valleys which contains a large momentum difference while the optical method are based on direct transitions. We propose a valley splitting at each band which can be measured via a two-terminal unipolar transport setup where a valley polarization is expected. Although specifying valley index is not as easy as spin index, we believe that the valley index can be realized through measuring spin resolved conductance in the TMDs due to the spin-valley coupling. In the unipolar case, the conductance can be calculated by counting the transport channel, so that the corresponding conductance for each spin component is given as $G^s_{nn (pp)}=min(\nu^s_L,\nu^s_R)$, in unit of $e^2/h$ between the left and right leads. In this regard, we plot the conductance based on the Landau level sequence of the two-band model in Fig.~\ref{fig:cond1} for both electron and hole doped cases and the spin polarization can also be seen. In the valence band the polarization is more pronounced due to the strong spin-orbit coupling. The sequence of the plateaus for both the cases are different in the low-energy levels. This effect can be understood based on the strong spin-orbit coupling in the valence band which decreases the number of the channel of the hole doped system to the half of the accessible channel in the conduction band.

\begin{figure}
\includegraphics[width=0.95\linewidth]{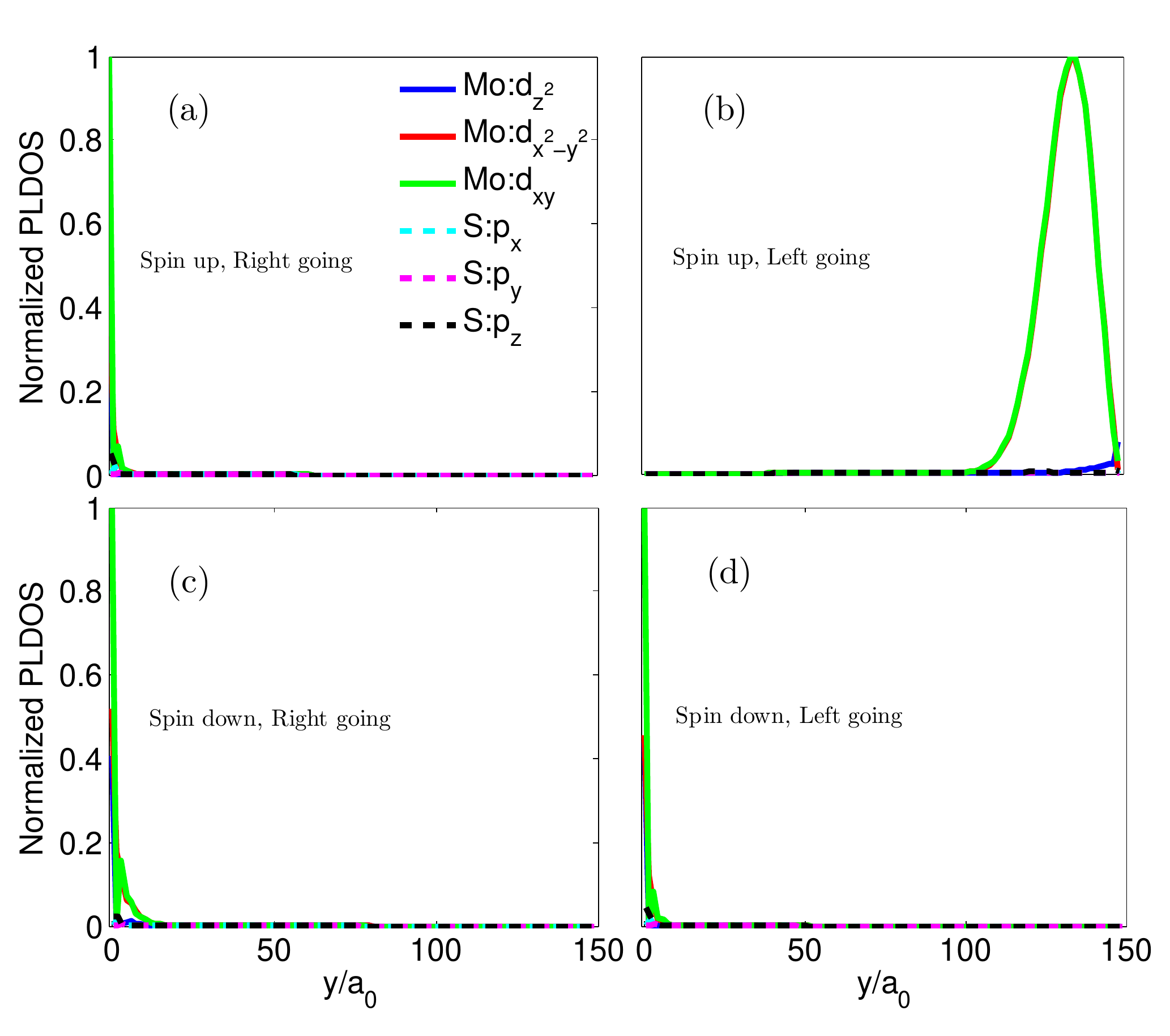}
\caption{(Color online) (a,b) Projected local density of states, $\rho(y, E_{k})$ for spin up edge modes at $E_K=-0.89$eV. The left and right going modes are localized on opposite edges. (c,d) the same as before for spin down edge modes but the left and right going states
are localized on same edges. The edge modes are mostly constructed by $d_{xy}$ and $d_{x^2-y^2}$ orbitals of the Molybdenum atoms.}
\label{fig:wave_up}
\end{figure}

Moreover, there are some finite size metallic edge modes ( see Fig.~\ref{fig:LL2}) due to the zigzag edges. These edge modes suppress the spin polarization when the system is subjected to an external magnetic field. We calculate the normalized projected local density of states (PLDOS) to clarify that each of those states are mostly localized on which edge and orbital. The PLDOS which can be calculated as $\rho(y,n,k,\mu)=\sum_{m k'}|\psi_{mk'\mu}(y)|^2\delta(E_{nk}-E_{mk'})$ is shown in Fig.~\ref{fig:wave_up} for spin up (a,d) and spin down (c,d) components, respectively. Here $\psi_{mk,\mu}$ is the wave function in which $ m (n)$, $k (k')$ and $\mu$ stand for the band index, momentum and orbital index, respectively. The left-going (which is defined by a negative slope of the dispensation relation) spin-up state, which is connected to the zero Landau level in the valence band at the $K$-point, lies on the top edge while the right-going one is located on the bottom edge. On the other hand, both right- and left- going spin-down states are on the bottom edge. This feature tells us that the former pair is chiral whereas the later one is not.

The non-equilibrium Green's function method is used in a two-terminal setup to count the number of the transport channel of a zigzag ribbon geometry. First of all, we calculate the conductance of a clean system in the presence of the external magnetic field and the results are illustrated in Fig.~\ref{fig:cond1} which shows the two-terminal conductance plateaus for each spin component. Obviously, there is no the spin-polarization for the low hole doped case and it is due to the extra finite size edge modes.

\begin{figure}
\includegraphics[width=0.95\linewidth]{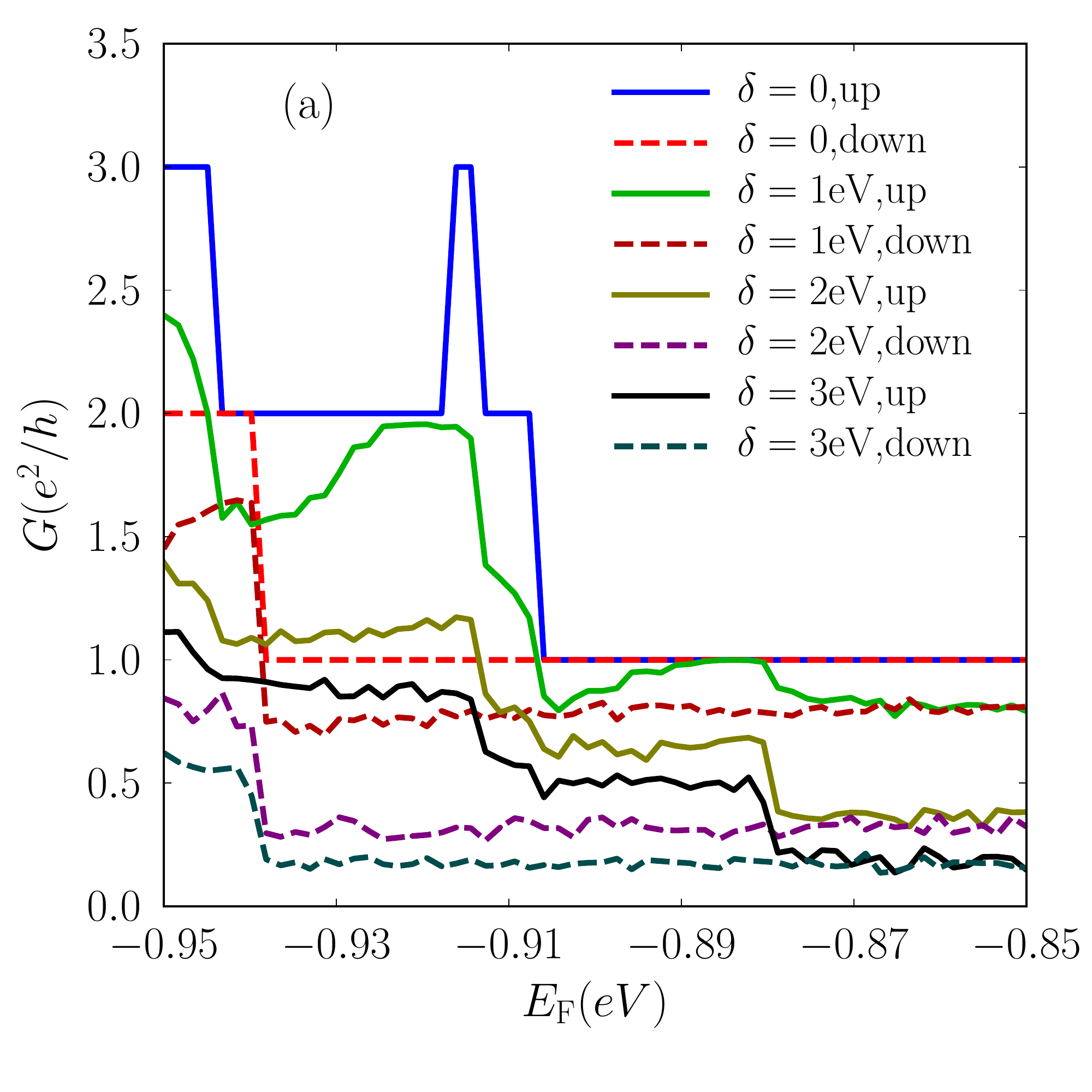}
\includegraphics[width=0.95\linewidth]{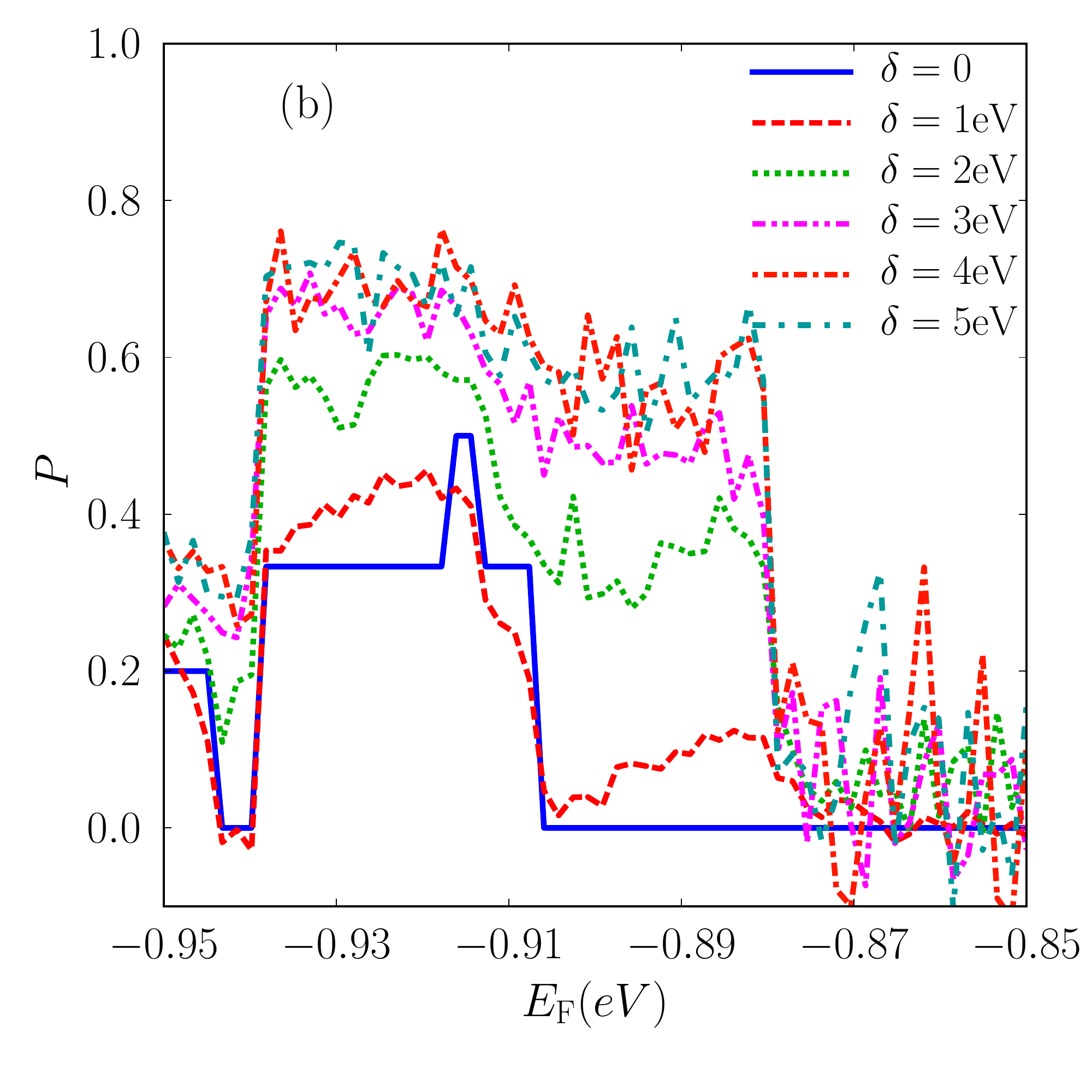}
\caption{(Color online) (a) Unipolar conductance for a zigzag ribbon as a function of the Fermi energy in the presence of the perpendicular magnetic field and random on-site energy.
(b) Spin polarization in the presence of the perpendicular magnetic field and random on-site energy. The Zeeman interaction corresponding to the real spin is not taken into account in this figure. We set $N=50$, $M=10$ and $B=150 T$.}
\label{fig:cond2}
\end{figure}

Furthermore, in a real material there are also impurities and structural defects which can affect the expected transport properties of the clean sample. Here, we study the effect of impurities by adding a simple random on-site energy in the range of $[-\delta/2,\delta/2]$ to the Hamiltonian where $\delta$ stands for the intensity of disorder scattering. In this case, we assume that all of the relevant atomic orbitals at each lattice site are affected in a same way from the presence of impurity. This kind of impurity which has a uniform distribution only induces an intra-valley scattering rate to relax the momentum. We are only interested in a simple momentum relaxation to realize whether finite size or quantum Hall edge modes are robust with respect to the randomness. The numerical conductance results as a function of the Fermi energy are presented in Fig.~\ref{fig:cond2} showing that disorder induces a spin-valley polarization. In the clean ribbon with low hole doped case, both spin components have same contributions to the conductance. The spin-down contribution of the conductance in the lowest plateau is originating from the finite size edge modes while that corresponding to the spin-up component has a contribution from a quantum Hall edge mode which is connected to the zero Landau level at the $K$-point.

After adding random on-site energy, one can clearly see that for a reasonable intensity of the randomness the spin-down edge modes are localized. This is due to the fact they are not chiral and thus they can scatter backward similar to a non-chiral one-dimensional system where a localization always occurs in the presence of a randomness. However, in the case of the spin-up sates, since they are on the opposite side of the ribbon, they can not be scattered to each other based on their chiral nature. Hence, the spin-up states are not localized and they can carry spin-polarized current which is also valley-polarized due to the spin-valley coupling of the hole doped case. Eventually, disorder revives the spin-valley polarized transport in the finite size case. Moreover, if we increase the strength of the scattering from impurity, the conductance contribution from both spin will drop, however the polarization will approximately saturate to a constant value ($P\sim0.6$).

\section{conclusion}
In this work, we have shown that the strength of the valley Zeeman interaction in TMDCs, which mainly originates from the broken inversion symmetry, differs in the conduction and valence bands due to the different orbital character and also virtual interband transitions. We have provided a modified two-band Hamiltonian in the presence of the magnetic field which can be used to describe recent experimental data. Moreover, we have shown that the quadratic diagonal momentum dependent terms in the low-energy model contribute in the valley splitting which evolves in a quadratic way by varying $B$ that might compete with the diamagnetic shift of the exciton binding energy. Remarkably, the dominant dependance of the valley splitting to the magnetic field, which evolves linearly with $B$, originates from the multi-orbital and multi-band structures of the system.

Furthermore, we have studied the two-terminal electronic transport of a zigzag ML-MoS$_2$ in the presence of a perpendicular magnetic field using the non-equilibrium recursive Green's function method. We have found that the conductance is not spin-polarized in the clean hole-doped case due to the presence of the finite size metallic edge modes in addition to the quantum Hall edge modes. Our numerical results in the two-terminal conductance show a spin-valley polarized transport in the presence of the on-site disorder which is related to the chiral nature of one of the spin components.

\begin{acknowledgments}
 We would like to thank F. Guinea for valuable discussions.
\end{acknowledgments}

\appendix
\section{Hopping matrices}
The hopping terms of the system, calculated by Slater-Koster table \cite{Gomez13}, are listed below for the nearest neighbor hopping,
\begin{widetext}
\begin{align}
t^{ab}_1&=\frac{\sqrt{2}}{7\sqrt{7}}\begin{pmatrix}-9 V_{pd\pi}+\sqrt{3}V_{pd\sigma}&&3\sqrt{3}V_{pd\pi}-V_{pd\sigma}&&12 V_{pd\pi}+\sqrt{3} V_{pd\sigma}\\
          5\sqrt{3} V_{pd\pi}+3 V_{pd\sigma}&&9 V_{pd\pi}-\sqrt{3} V_{pd\sigma}&&-2\sqrt{3}V_{pd\pi}+3 V_{pd\sigma}\\
          -V_{pd\pi}-3\sqrt{3}V_{pd\sigma}&&5\sqrt{3}V_{pd\pi}+3 V_{pd\sigma}&&6 V_{pd\pi}-3\sqrt{3} V_{pd\sigma}\end{pmatrix}\\
\nonumber\\
t^{ab}_2&=\frac{\sqrt{2}}{7\sqrt{7}}\begin{pmatrix}0&&-6\sqrt{3}V_{pd\pi}+2V_{pd\sigma}&&12V_{pd\pi}+\sqrt{3}V_{pd\sigma}\\
          0&&-6V_{pd\pi}-4\sqrt{3}V_{pd\sigma}&&4\sqrt{3} V_{pd\pi}-6V_{pd\sigma}\\14V_{pd\pi}&&0&&0\end{pmatrix}\\
\nonumber\\
t^{ab}_3&=\frac{\sqrt{2}}{7\sqrt{7}}\begin{pmatrix}9 V_{pd\pi}-\sqrt{3}V_{pd\sigma}&&3\sqrt{3}V_{pd\pi}-V_{pd\sigma}&&12 V_{pd\pi}+\sqrt{3} V_{pd\sigma}\\
          -5\sqrt{3} V_{pd\pi}-3 V_{pd\sigma}&&9 V_{pd\pi}-\sqrt{3} V_{pd\sigma}&&-2\sqrt{3}V_{pd\pi}+3 V_{pd\sigma}\\
          -V_{pd\pi}-3\sqrt{3}V_{pd\sigma}&&-5\sqrt{3}V_{pd\pi}-3 V_{pd\sigma}&&-6V_{pd\pi}+3\sqrt{3} V_{pd\sigma}\end{pmatrix}
\end{align}
\end{widetext}
The next nearest neighbor hopping process, the hopping along $a_i$ direction (see Fig.\ref{scheme}) which corresponds to the hopping among the Mo or the S atoms, reads as
\begin{widetext}
\begin{align}
t^{aa}_1&=\frac{1}{4}\begin{pmatrix}3V_{dd\delta}+V_{dd\sigma}&&\frac{\sqrt{3}}{2}(-V_{dd\delta}+V_{dd\sigma})&&-\frac{3}{2}(V_{dd\delta}-V_{dd\sigma})\\
        \frac{\sqrt{3}}{2}(-V_{dd\delta}+V_{dd\sigma})&&\frac{1}{4}(V_{dd\delta}+12V_{dd\pi}+3V_{dd\sigma})&&\frac{\sqrt{3}}{4}(V_{dd\delta}-4V_{dd\pi}+3V_{dd\sigma})\\
        -\frac{3}{2}(V_{dd\delta}-V_{dd\sigma})&&\frac{\sqrt{3}}{4}(V_{dd\delta}-4V_{dd\pi}+3V_{dd\sigma})&&\frac{1}{4}(3V_{dd\delta}+4V_{dd\pi}+9V_{dd\sigma})\end{pmatrix}\\
\nonumber\\
t^{aa}_2&=\frac{1}{4}\begin{pmatrix}3V_{dd\delta}+V_{dd\sigma}&&\sqrt{3}(V_{dd\delta}-V_{dd\sigma})&&0\\\sqrt{3}(V_{dd\delta}-V_{dd\sigma})&&V_{dd\delta}+3V_{dd\sigma}&&0\\0&&0&&4V_{dd\pi}\end{pmatrix}\\
\nonumber\\
t^{aa}_3&=\frac{1}{4}\begin{pmatrix}3V_{dd\delta}+V_{dd\sigma}&&\frac{\sqrt{3}}{2}(-V_{dd\delta}+V_{dd\sigma})&&\frac{3}{2}(V_{dd\delta}-V_{dd\sigma})\\
        \frac{\sqrt{3}}{2}(-V_{dd\delta}+V_{dd\sigma})&&\frac{1}{4}(V_{dd\delta}+12V_{dd\pi}+3V_{dd\sigma})&&-\frac{\sqrt{3}}{4}(V_{dd\delta}-4V_{dd\pi}+3V_{dd\sigma})\\
        \frac{3}{2}(V_{dd\delta}-V_{dd\sigma})&&-\frac{\sqrt{3}}{4}(V_{dd\delta}-4V_{dd\pi}+3V_{dd\sigma})&&\frac{1}{4}(3V_{dd\delta}+4V_{dd\pi}+9V_{dd\sigma})\end{pmatrix}\\
\nonumber\\
t^{bb}_1&=\frac{1}{4}\begin{pmatrix}3V_{pp\pi}+V_{pp\sigma}&&\sqrt{3}(V_{pp\pi}-V_{pp\sigma})&&0\\
                                     \sqrt{3}(V_{pp\pi}-V_{pp\sigma})&&V_{pp\pi}+3V_{pp\sigma}&&
                                     0\\0&&0&&4V_{pp\pi}\end{pmatrix}\\
\nonumber\\
t^{bb}_2&=\begin{pmatrix}V_{pp\sigma}&&0&&0\\0&&V_{pp\pi}&&0\\0&&0&&V_{pp\pi}\end{pmatrix}\\
\nonumber\\
t^{bb}_3&=\frac{1}{4}\begin{pmatrix}3V_{pp\pi}+V_{pp\sigma}&&-\sqrt{3}(V_{pp\pi}-V_{pp\sigma})&&0\\-\sqrt{3}(V_{pp\pi}-V_{pp\sigma})&&V_{pp\pi}+3V_{pp\sigma}
                                      &&0\\0&&0&&4V_{pp\pi}\end{pmatrix}
\end{align}
\end{widetext}

The direction of the hopping indicated by subindex 1,2, and 3 can be seen in Fig.~\ref{scheme} for the nearest and next nearest neighbor hopping. Note that $a=\sqrt{3}a_0=0.316$nm stands for the Mo-Mo or in plane S-S bond length with $a_0$ as in plane projection of the Mo-S bond length.

\end{document}